\documentclass[fleqn,twoside]{article}
\usepackage{espcrc2}

\usepackage{graphicx}
\usepackage[figuresright]{rotating}

\newcommand{\AmS}{{\protect\the\textfont2
  A\kern-.1667em\lower.5ex\hbox{M}\kern-.125emS}}

\newcommand{\be}{\begin{equation}}
\newcommand{\bea}{\begin{eqnarray}}
\newcommand{\ee}{\end{equation}}
\newcommand{\eea}{\end{eqnarray}}
\newcommand{\bpi}{\begin{picture}}
\newcommand{\bce}{\begin{center}}
\newcommand{\epi}{\end{picture}}
\newcommand{\ece}{\end{center}}

\def\chic#1{{\scriptscriptstyle #1}}

\def\gb{{\rm I}\hspace{-0.07cm}\Gamma}

\def\gb{{\rm I}\hspace{-0.07cm}\Gamma}

\title{Non-perturbative QCD effective charges}

\author{Arlene~C.~Aguilar \address{Federal University of ABC, CCNH,\\ 
Rua Santa Ad\'elia 166,  CEP 09210-170, Santo Andr\'e, Brazil} 
and Joannis~Papavassiliou \address{Department of Theoretical Physics and IFIC, \\
University of Valencia-CSIC,
E-46100, Valencia, Spain}}%
              
\begin{document}

\begin{abstract}

Using  gluon  and ghost  propagators obtained
from Schwinger-Dyson  equations (SDEs),
we construct the non-perturbative effective charge of QCD.  We employ two
different definitions,  which, despite their  distinct field-theoretic
origin, give  rise to qualitative  comparable results, by virtue  of a
crucial non-perturbative identity.   Most importantly,  the  QCD charge
obtained  with either  definition  freezes in  the  deep infrared,  in
agreement  with theoretical  and  phenomenological expectations.   The
various  theoretical ingredients necessary  for this  construction are
reviewed  in  detail,  and  some conceptual subtleties are  briefly
discussed.

\vspace{1pc}
\end{abstract}

\maketitle

\section{Introduction}

One of the challenges of the QCD is the self-consistent 
and physically meaningful definition of an effective charge. 
This quantity provides a continuous 
interpolation between two physically distinct regimes: the deep ultraviolet (UV), where 
perturbation theory works well, and the deep infrared (IR), where 
non-perturbative techniques, such as lattice or SDEs, must be employed. 
The effective charge depends strongly 
on the detailed dynamics of some of the most fundamental 
Green's functions of QCD, such as the gluon and ghost propagators~\cite{Aguilar:2009nf}.

In this talk we will focus on  
two characteristic definitions of the effective charge,  
frequently employed in the literature. Specifically we will 
consider (i) the effective charge of the pinch technique (PT)~\cite{Cornwall:1982zr,Nair:2005iw} 
and (ii) the one obtained from the ghost-gluon vertex~\cite{Alkofer:2004it}.

\section{\label{seffch} Two non-perturbative effective charges}

We first introduce  the notation and the basic quantities entering into our study. 
In the covariant  renormalizable ($R_\xi$) gauges, the gluon
propagator $\Delta_{\mu\nu}(q)$  has the form  
\be \Delta_{\mu\nu}(q)=-i\left[ P_{\mu\nu}(q)\Delta(q^2) +\xi\frac{q_\mu q_\nu}{q^4}\right],
\label{prop_cov}
\ee
where $\xi$ denotes the gauge-fixing parameter,
\mbox{$P_{\mu\nu}(q)= g_{\mu\nu} - q_\mu q_\nu /q^2$}
is the usual transverse projector, and \mbox{$\Delta^{-1}(q^2) = q^2 + i \Pi(q^2)$}, 
with  \mbox{$\Pi_{\mu\nu}(q)=P_{\mu\nu}(q)\Pi(q^2)$} the gluon self-energy.  
In addition, the full ghost propagator $D(q^2)$ and its dressing function $F(q^2)$
are related by \mbox{$iF(q^2)=q^2 D(q^2)$.}
The all-order ghost vertex 
will be denoted by  $\gb_\mu(k,q)$, with $k$ representing the momentum 
of the gluon and $q$ the one of the anti-ghost;
at tree-level $\gb^{(0)}_\mu(k,q)= -q_\mu$.

\begin{figure}[!t]
\begin{center}
\includegraphics[scale=0.45]{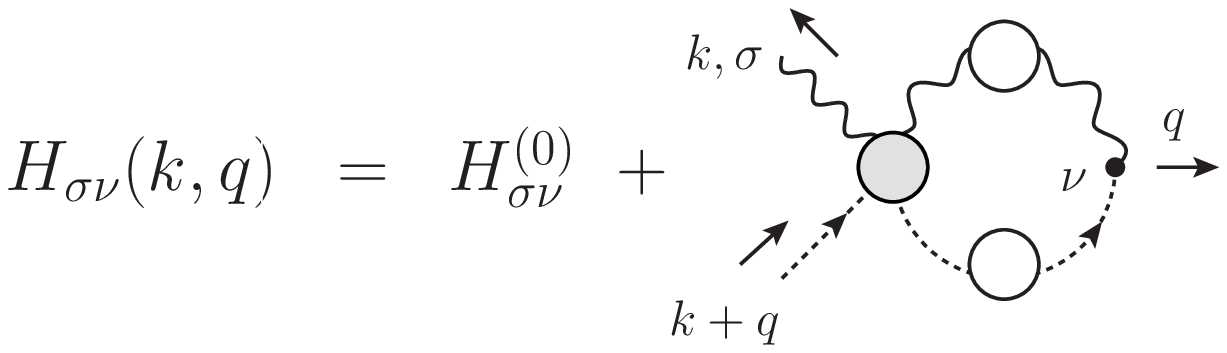}
\end{center}
\vspace{-1.5cm}
\caption{Diagrammatic representation of $H$.}
\label{fig:Lambda_aux}
\vspace{-0.5cm}
\end{figure}

An important ingredient for what follows is 
the two-point function $\Lambda_{\mu\nu}(q)$ defined by \cite{Aguilar:2009nf,Aguilar:2008xm}
\bea
i \Lambda_{\mu \nu}(q) &=& \lambda
\int_k H^{(0)}_{\mu\rho}
D(k+q)\Delta^{\rho\sigma}(k)\, H_{\sigma\nu}(k,q),
\nonumber \\
&=& g_{\mu\nu} G(q^2) + \frac{q_{\mu}q_{\nu}}{q^2} L(q^2),
\label{LDec}
\eea
where \mbox{$\lambda=g^2C_{\rm {A}}$},  $C_{\rm {A}}$ is the Casimir eigenvalue of the adjoint representation, 
and \mbox{$\int_{k}\equiv\mu^{2\varepsilon}(2\pi)^{-d}\int\!d^d k$}, 
with $d$ the dimension of space-time. 
$H_{\mu\nu}(k,q)$ is represented in Fig.~\ref{fig:Lambda_aux},  and its tree-level, \mbox{$H_{\mu\nu}^{(0)} = ig_{\mu\nu}$}. In addition,  $H_{\mu\nu}(k,q)$ 
is related to $\gb_\mu(k,q)$ by
\be
q^\nu H_{\mu\nu}(k,q)=-i\gb_{\mu}(k,q)\,.
\label{qH}
\ee 

\subsection{The pinch technique effective charge}

\begin{figure}[!b]
\begin{center}
\includegraphics[scale=0.4]{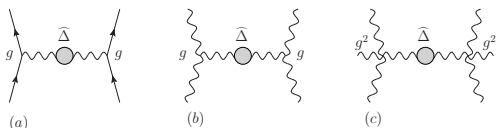}
\end{center}
\vspace{-1.5cm}
\caption{The universal PT coupling.}
\label{fig:pt_coup}
\vspace{-1.5cm}
\end{figure}

The PT definition of the effective charge 
relies on the construction of an universal ({\it i.e.}, process-independent) 
effective gluon propagator,  
which captures the running of the QCD $\beta$ function, exactly as happens with the 
vacuum polarization in the case of QED \cite{Cornwall:1982zr,Watson:1996fg} (See Fig.~\ref{fig:pt_coup}).
One important point, explained in detail in the literature, is the (all-order) correspondence 
between the PT and the Feynman gauge of the BFM~
\cite{Cornwall:1982zr,Abbott:1980hw}. 
In fact, one can generalize the PT construction \cite{Cornwall:1982zr}
in such a way as to reach diagrammatically 
any value of the gauge fixing parameter of the BFM, and in particular the Landau gauge.
In what follows we will implicitly assume the aforementioned generalization of the PT, 
given that the main identity we will use to 
relate the two effective charges is valid only in the Landau gauge.

To fix the ideas,  the PT one-loop gluon self-energy reads
\be 
\widehat\Delta^{-1}(q^2)= q^2\left[1+ b g^2\ln\left(\frac{q^2}{\mu^2}\right)\right],
\label{rightRG}
\ee
where  $b = 11 C_A/48\pi^2$  is the first coefficient of the QCD $\beta$-function. 
Due to the Abelian WIs satisfied by the PT effective Green's functions, the 
 new propagator-like quantity $\widehat\Delta^{-1}(q^2)$ absorbs all  
the RG-logs, exactly as happens in QED with the photon self-energy.
Then, the renormalization 
constants of the gauge-coupling and of the PT gluon self-energy, 
defined as 
\bea
g(\mu^2) &=&Z_g^{-1}(\mu^2) g_0 ,\nonumber \\
\widehat\Delta(q^2,\mu^2) & = & \widehat{Z}^{-1}_A(\mu^2)\widehat{\Delta}_0(q^2), 
\label{conrendef}
\eea
where the ``0'' subscript indicates bare quantities, satisfy the 
QED-like relation \mbox{${Z}_{g} = {\widehat Z}^{-1/2}_{A}$}.
Therefore, the product 
\be
{\widehat d}_0(q^2) = g^2_0 \widehat\Delta_0(q^2) = 
g^2(\mu^2) \widehat\Delta(q^2,\mu^2) = {\widehat d}(q^2), 
\label{ptrgi}
\ee
forms a RG-invariant ($\mu$-independent) quantity~\cite{Cornwall:1982zr}.
For asymptotically large momenta one may extract from ${\widehat d}(q^2)$
a dimensionless quantity by writing,
\be
{\widehat d}(q^2) = \frac{\overline{g}^2(q^2)}{q^2},
\label{ddef1}
\ee
where $\overline{g}^2(q^2)$ is the RG-invariant effective charge of QCD; at one-loop
\be
\overline{g}^2(q^2) = \frac{g^2}{1+  b g^2\ln\left(q^2/\mu^2\right)}
= \frac{1}{b\ln\left(q^2/\Lambda^2_\chic{\mathrm{QCD}}\right)}.
\label{effch}
\ee
where $\Lambda_\chic{\mathrm{QCD}}$ denotes an RG-invariant mass scale of a few hundred ${\rm MeV}$.

Eq.~(\ref{ptrgi}) is a non-perturbative relation; therefore it can serve unaltered as the starting 
point for extracting a non-perturbative effective charge, 
provided that one has information on the IR behavior of the PT-BFM gluon propagator $\widehat\Delta(q^2)$.
Interestingly enough, non-perturbative information on the {\it conventional} gluon propagator $\Delta(q^2)$ 
may also be used, by virtue of a general relation connecting $\Delta(q^2)$ and $\widehat\Delta(q^2)$.
Specifically, a formal all-order relation known as ``background-quantum'' identity ~\cite{Grassi:1999tp}
states that 
\be
\Delta(q^2) = 
\left[1+G(q^2)\right]^2 \widehat{\Delta}(q^2).
\label{bqi2}
\ee
Note that, due to its BRST origin, the above relation must be preserved after renormalization. 
Specifically, denoting by $Z_\Lambda$ the renormalization constant relating 
the bare and renormalized functions, $\Lambda_0^{\mu\nu}$ and $\Lambda^{\mu\nu}$, through
\be
g^{\mu\nu} + \Lambda^{\mu\nu}(q,\mu^2)=Z_\Lambda(\mu^2)[g^{\mu\nu}+ \Lambda_0^{\mu\nu}(q)],
\label{Lamrel}
\ee
then from (\ref{bqi2}) and \mbox{${Z}_{g} = {\widehat Z}^{-1/2}_{A}$} follows the additional relation 
\be
Z_g^{-1} = Z_A^{1/2} Z_\Lambda ,
\label{extrel}
\ee 
which is useful for the comparison with the coupling discussed 
in the following subsection.

It is now easy to verify, at lowest order, that 
the $1+G(q^2)$ obtained from Eq.~(\ref{LDec})  restores the $\beta$ function coefficient  
in front of UV logarithm. In that limit~\cite{Aguilar:2008xm}
\bea
1+G(q^2) &=& 1 +\frac{9}{4}
\frac{C_{\rm {A}}g^2}{48\pi^2}\ln\left(\frac{q^2}{\mu^2}\right),\nonumber \\
\Delta^{-1}(q^2) &=& q^2 \left[1+\frac{13}{2}
\frac{C_{\rm {A}}g^2}{48\pi^2}\ln\left(\frac{q^2}{\mu^2}\right)\right].
\label{pert_gluon}
\eea
Using  Eq.~(\ref{bqi2}) we therefore recover the $\widehat{\Delta}^{-1}(q^2)$ 
of Eq.~(\ref{rightRG}), as we should. 

Then, non-perturbatively, one substitutes into  Eq.~(\ref{bqi2}) the $ 1+G(q^2)$ and $\Delta(q^2)$ 
obtained from either the lattice or SD analysis, to obtain $\widehat{\Delta}(q^2)$.  
This latter quantity is the non-perturbative generalization of Eq.~(\ref{rightRG}); 
for the same reasons explained above, the combination 
\be
\widehat{d}(q^2)= \frac{g^2 \Delta(q^2)}{\left[1+G(q^2)\right]^2}\,,
\label{rgi}
\ee
is an RG-invariant quantity.

\subsection{The ghost-gluon vertex charge}

In principle, a definition for the QCD effective charge can be obtained starting 
from the various QCD vertices; however, in general, such a construction 
involves more than one momentum scales, and further assumptions about their values need be introduced, in order to express the charge as a function of a single variable. 
The ghost-gluon vertex has been particularly popular in this context, 
especially in conjunction with Taylor's non-renormalization theorem and 
the corresponding kinematics~\cite{Alkofer:2004it}.

We next define  
the following renormalization constants
%
%
\bea
\Delta(q^2)= Z^{-1}_{A}\Delta_0(q^2), &&
F(q^2)= Z^{-1}_{c}F_0(q^2), \nonumber\\ 
\gb^\nu(k,q)= Z_1\gb^\nu_0(k,q), &&
g^{\prime} = Z_{g^{\prime}}^{-1} g_0\,.
\label{renconst}
\eea

Notice that a priori $Z_{g^{\prime}}$ defined as \mbox{$Z_{g'}= Z_1 Z_A^{-1/2} Z_c^{-1}$},
does not have to coincide with the 
$Z_{g}$ introduced in (\ref{conrendef}); however, as we will see 
shortly, they do coincide by virtue of the basic identity we will
derive in next section. 

It turns out that for the so-called Taylor  
kinematics (vanishing incoming ghost momentum, $k_\mu \to -q_\mu$), one may impose the additional condition 
\be
Z_1= Z_{g^{\prime}} Z_A^{1/2}Z_c=1 \;\;\Rightarrow\;\;\ Z_{g^{\prime}}^{-1} = Z_A^{1/2}Z_c\,.
\label{Z1T}
\ee
Thus, the combination 
\be
\widehat{r}(q^2) \ = \ {g'}^2 \Delta(q^2) F^2(q^2)\,,
\label{rg2}
\ee
is a RG-invariant ($\mu$-independent) quantity. 
Therefore, for asymptotically large $q^2$, in analogy to Eq.~(\ref{ddef1}) one can define  
an alternative QCD running coupling as 
\be
\widehat{r}(q^2)=\frac{\overline{g}_{\mathrm{gh}}^2(q^2)}{q^2}.
\ee 
It is easy to verify that  $\overline{g}^2_\mathrm{gh}(q^2)$ and  $\overline{g}^2(q^2)$ displays the same one-loop 
behavior, since, perturbatively, the function \mbox{$1+G(q^2)$} is the inverse of the ghost dressing function $F(q^2)$ 
(this is due to the general identity of Eq.~\ref{funrel}).

\section{\label{giSDE} An important identity}

In this section, we discuss a non-trivial identity, valid  {\it only} in the Landau gauge, 
relating the $F(q^2)$ with the $G(q^2)$ and $L(q^2)$ of (\ref{LDec}). 

\begin{figure}[tbp]
\begin{center}
\includegraphics[scale=0.5]{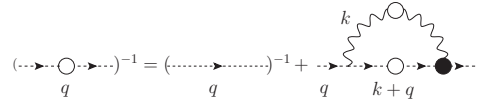}
\vspace{-1.0cm}
\caption{The SDE for the ghost propagator.}
\label{fig:ghost}
\vspace{-0.5cm}
\end{center}
\end{figure}

The derivation proceeds as follows.
First, consider the standard SD equation for the ghost propagator, represented
in Fig.~\ref{fig:ghost}, and written as
\be
iD^{-1}(q^2) = q^2 +i  \lambda \int_k
\Gamma^{\mu}\Delta_{\mu\nu}(k)\gb^{\nu}(k,q) D(p),
\label{SDgh}
\ee
where \mbox{$p=k+q$}. Then, contract both sides of
the defining equation (\ref{LDec}) by the combination $q^{\mu}q^{\nu}$ to get
\be
[G(q^2) + L(q^2)]q^2\!\!=\!\!\lambda\!\!
\int_k\!\! q_{\rho} \Delta^{\rho\sigma}(k)\, q^{\nu} H_{\sigma\nu}(k,q) D(p).
\label{s1}
\ee
Using the Eq.~(\ref{qH}) and the transversality of the full gluon propagator, we can see that the rhs of Eq.~(\ref{s1}) 
is precisely the integral appearing in the ghost SDE~(\ref{SDgh}). Therefore, in terms of the ghost dressing function $F(q^2)$, 
\be
1+ G(q^2) + L(q^2) = F^{-1}(q^2).
\label{funrel}
\ee
Eq.~(\ref{funrel}), derived here from the SDEs,  
has been first obtained in~\cite{Kugo:1995km},
as a direct consequence of the BRST symmetry.

Let us study the functions \mbox{$G(q^2)$} and  \mbox{$L(q^2)$} more closely. From Eq.~(\ref{LDec}) we have that (in $d$ dimensions) 
\bea
G(q^2)&=&\frac{1}{(d-1)q^2} \left(q^2 \Lambda_{\mu}^{\mu} - q^{\mu}q^{\nu}\Lambda_{\mu\nu} \right),
\nonumber\\
L(q^2)&=&\frac{1}{(d-1)q^2} \left(d q^{\mu}q^{\nu}\Lambda_{\mu\nu} - q^2\Lambda_{\mu}^{\mu} \right)\,.
\label{s3}
\eea 
In order to study the relevant equations further,
we will approximate the two vertices, \mbox{$H_{\mu\nu}$} and $\gb_\mu$,
by their tree-level values, 
Then, setting \mbox{$f(k,q) \equiv  (k \cdot q)^2/{k^2 q^2}$},
one may show that \cite{Aguilar:2009nf} 
\bea
F^{-1}(q^2)\!\!\!\!\!&=&\!\!\!\!\!1 \,\,+\,\, \lambda\!\int_k \!\![1\!\!-\!\!f(k,q)] \Delta (k)  D(k+q),
\nonumber\\
(d-1) G(q^2)\!\!\!\!\!&=&\!\!\!\!\! \lambda\!\int_k\!\![
(d-2)\!\!+\!\! f(k,q)]\Delta (k)  D(q+k),
\nonumber\\
(d-1) L(q^2)\!\!\!\!\!&=&\!\!\!\!\!\lambda\!\int_k\!\![1\!\!-\!\!d \,f(k,q)]\Delta (k)  D(q+k)\,,
\label{simple}
\eea 
It turns out that if $F$ and $\Delta$ are both IR finite,   
Eq.~(\ref{simple}) yields the important result 
$L(0)= 0$ \cite{Aguilar:2009nf}. 

Of course, all quantities appearing in Eq.~(\ref{simple}) are unrenormalized. 
It is easy to recognize, for example,  by substituting in the corresponding 
integrals tree-level expressions,
that  $F^{-1}(q^2)$ and  $G(q^2)$ have exactly the same 
logarithmic dependence on the UV cutoff,  
while $L(q^2)$ is finite at leading order.

Since the origin of (\ref{funrel}) 
is the BRST symmetry,  it should not be deformed after renormalization.
To that end, using the definition of (\ref{Lamrel}), 
in order to preserve the relation~(\ref{funrel}) we must impose that 
\mbox{$Z_\Lambda = Z_{c}$}.
In addition, by virtue of (\ref{qH}), and for the same reason,  
we have that, in the Landau gauge, ${\gb}_{\nu}(k,q)$ and $H_{\sigma\nu}(k,q)$ 
must be renormalized by the same renormalization constant, namely $Z_1$ 
[{\it viz.} Eq.~(\ref{renconst})]; for the Taylor kinematics, we have that $Z_1=1$ 
[see Eq.~(\ref{Z1T})]  (for some additional subtleties see \cite{Aguilar:2009nf}).

Returning to the effective charges, 
first of all, comparing Eq.~(\ref{ptrgi}) and Eq.~(\ref{rg2}), it is clear that $g(\mu)=g^{\prime}(\mu)$,
by virtue of  \mbox{$Z_\Lambda = Z_{c}$}. Therefore, using Eq.~(\ref{bqi2}), one can get a relation between the two RG-invariant quantities, 
$\widehat{r}(q^2)$ and $\widehat d(q^2)$, namely
\be
\widehat{r}(q^2)=[1+G(q^2)]^2 F^2(q^2)\widehat d(q^2).
\label{rel_rgi}
\ee
From this last equality follows that $\alpha_{\chic{\mathrm{PT}}}(q^2)$ and $\alpha_{\mathrm{gh}}(q^2)$ 
are related by 
\be
\alpha_{\mathrm{gh}}(q^2) = [1+G(q^2)]^2 F^2(q^2)\alpha_{\chic{\mathrm{PT}}}(q^2)\,.
\label{coup_charge}
\ee
After using Eq.~(\ref{funrel}), we have that  
\be
\alpha_{\chic{\mathrm{PT}}}(q^2) = \alpha_{\mathrm{gh}}(q^2)\left[1+ \frac{L(q^2)}{1+G(q^2)}\right]^2 \,.
\label{relcoup2}
\ee
Evidently, the two couplings can only coincide at two points:
(i) at $q^2=0$, where, 
due to the fact that $L(0)=0$, 
we have that 
$\alpha_{\mathrm{gh}}(0) = \alpha_{\chic{\mathrm{PT}}}(0)$, 
and (ii) in the deep UV, where $L(q^2)$ approaches a constant.

\section{The nonperturbative analysis}

We next turn to the dynamical information required for the various ingredients entering into 
the effective charges defined above. To that end, we 
solve numerically the system of SDEs for $\Delta(q^2)$ 
$F(q^2)$, $G(q^2)$  and $L(q^2)$ obtained in~\cite{Aguilar:2008xm} 

\begin{figure}[t]
\begin{center}
\includegraphics[scale=0.7]{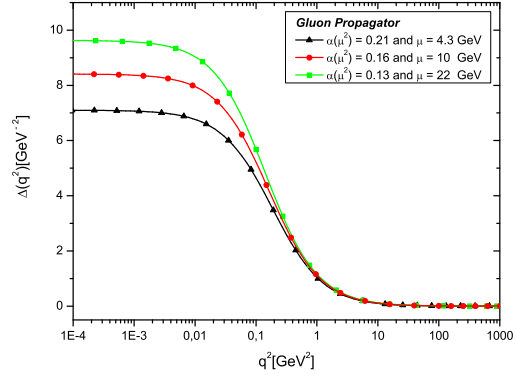}
\vspace{-1.7 cm}
\caption{Numerical solutions for the gluon propagator.}
\label{fig:plot1}
\vspace{-1.0cm}
\end{center}
\end{figure}

\begin{figure}[t]
\begin{center}
\includegraphics[scale=0.7]{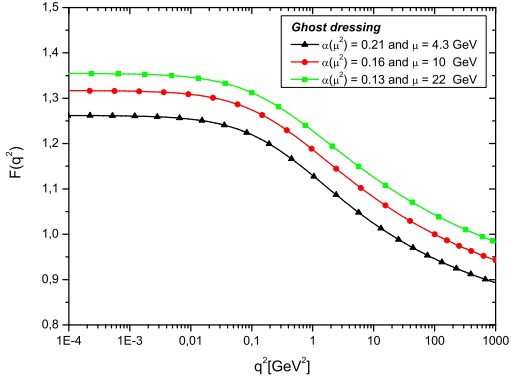}
\vspace{-1.7cm}
\caption{Numerical solutions for the ghost dressing function.}
\label{fig:plot2}
\vspace{-1.0cm}
\end{center}
\end{figure}

In Figs.~(\ref{fig:plot1}) and (\ref{fig:plot2}) we show the results for $\Delta(q^2)$ and $F(q^2)$ renormalized at three different points, \mbox{$\mu = \{4.3, 10, 22\}$} \mbox{GeV} with
  \mbox{$\alpha(\mu^2)=\{0.21, 0.16,0.13\}$} respectively. On the right panel we plot the corresponding $F(q^2)$ renormalized at the same points.  Notice that the solutions obtained are in qualitative agreement with recent results from large-volume lattices \cite{Cucchieri:2007md} where the both quantities, $\Delta(q^2)$ and $Fq^2)$, reach finite (non-vanishing) values in the deep IR.   

\begin{figure}[!b]
\begin{center}
\includegraphics[scale=0.7]{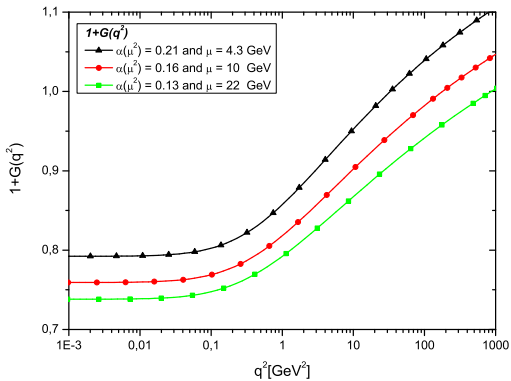}
\vspace{-1.7cm}
\caption{$1+G(q^2)$ determined from Eq.~(\ref{simple}).}
\label{fig:plot3}
\vspace{-1.5cm}
\end{center}
\end{figure}

\begin{figure}[!t]
\begin{center}
\includegraphics[scale=0.7]{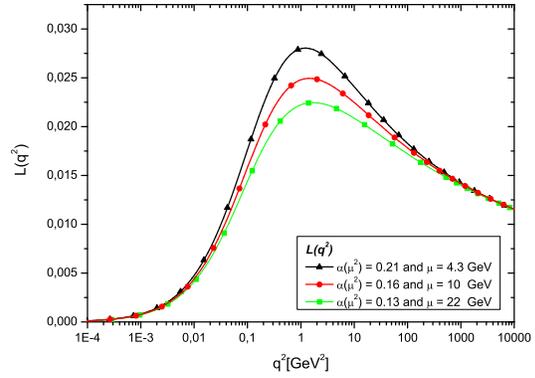}
\vspace{-1.7cm}
\caption{$L(q^2)$ determined from Eq.~(\ref{simple})}
\label{fig:plot4}
\vspace{-1.5cm}
\end{center}
\end{figure}

The results  for $1+G(q^2)$ and $L(q^2)$, renormalized at the same points, are presented in Fig.~\ref{fig:plot3}. As we can see, the function $1+G(q^2)$ is also IR finite exhibiting a plateau for  values of $q^2<0.1 \mbox{GeV}^2$. In the UV region, we instead recover the perturbative behavior~(\ref{pert_gluon}). On the other hand, $L(q^2)$, Fig.~\ref{fig:plot4},  shows a maximum in the intermediate momentum region, while, as expected, $L(0)=0$.


\begin{figure}[!t]
\begin{center}
\includegraphics[scale=0.7]{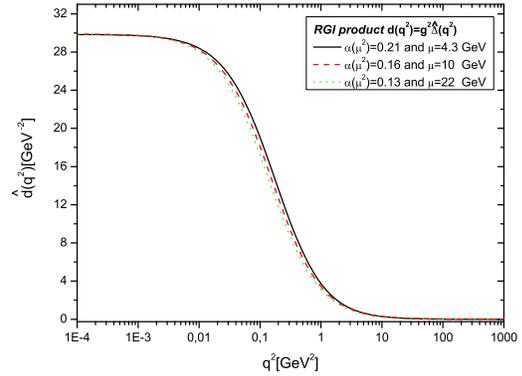}
\vspace{-1.5cm}
\caption{The $\widehat{d}(q^2)$ obtained by combining $\Delta(q^2)$ and $[1+G(q^2)]^2$according to (\ref{rgi}).}
\label{fig:plot5}
\vspace{-1.0cm}
\end{center}
\end{figure}

With all ingredients defined, the first thing one can check is that indeed Eq.~(\ref{rgi}) 
gives rise to a RG-invariant combination. Using the latter definition,  we can combine  
the different data sets  for $\Delta(q^2)$ and $[1+G(q^2)]^2$ at different renormalization points, to arrive at  the curves shown in Fig.~\ref{fig:plot5}.  Indeed, we see that all curves, for different values of $\mu$, merge one into the other proving 
that the combination $\widehat{d}(q^2)$ is independent of the renormalization point chosen.

From the dimensionful ${\widehat d}(q^2)$ we must now extract 
a dimensionless factor, ${\overline g}^2(q^2)$, corresponding to the running coupling (effective charge).  
Given that $\Delta(q^2)$ is IR finite (no more ``scaling''!), 
the physically meaningful procedure is to factor out from 
${\widehat d}(q^2)$ a massive propagator $[q^2+m^2(q^2)]^{-1}$, 
\be
\widehat{d}(q^2) = \frac{\overline{g}^2(q^2)}{q^2 + m^2(q^2)}\,,
\label{ddef}
\ee
where for the mass we will assume  ``power-law'' running~\cite{Lavelle:1991ve}, 
$m^2(q^2)= m^4_0/(q^2+m^2_0)$.

\begin{figure}[!t]
\begin{center}
\includegraphics[scale=0.7]{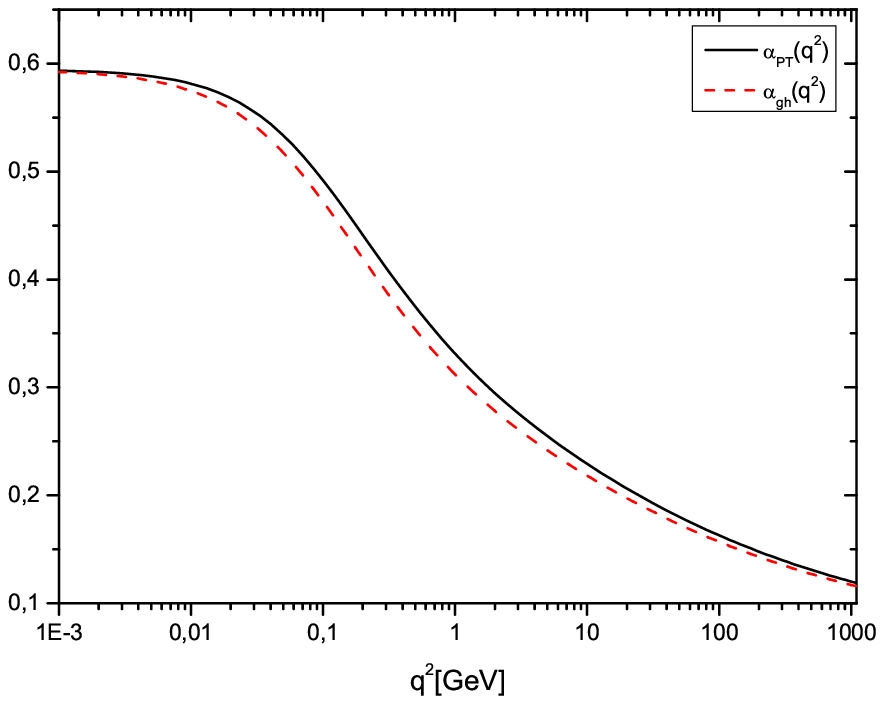}
\vspace{-1.5cm}
\caption{$\alpha_{\mathrm{gh}}(q^2)$ vs $\alpha_{\mathrm{PT}}(q^2)$, for \mbox{$m_0=500 \,\mbox{MeV}$}. }
\label{fig:plot6}
\vspace{-1.2cm}
\end{center}
\end{figure}

Thus, it follows from Eq.~(\ref{ddef}), that the effective charge $\alpha_{\chic{\mathrm{PT}}} = \overline{g}^2(q^2)/4\pi$ is identified as being
\be
4\pi\alpha_{\chic{\mathrm{PT}}}(q^2)= [q^2+m^2(q^2)]\widehat{d}(q^2),
\label{charge}
\ee 

Finally we compare numerically the two effective charges, $\alpha_{\chic{\mathrm{PT}}}(q^2)$ and $\alpha_{\mathrm{gh}}(q^2)$ in Fig.~\ref{fig:plot6}. First, we determine $\alpha_{\chic{\mathrm{PT}}}(q^2)$ obtained using (\ref{charge}), then we obtain $\alpha_{\mathrm{gh}}(q^2)$ with help of (\ref{relcoup2}) and the results for $1+G(q^2)$ and $L(q^2)$, Fig.~\ref{fig:plot3} and  Fig.~\ref{fig:plot4}.

As we can clearly see, both couplings freeze  at the same  finite value,
exhibiting a plateau for values of $q^2<0.02\, \mbox{GeV}^2$,  
while in the UV both show the expected  perturbative behavior. They differ only slightly in the intermediate region where
the values of $L(q^2)$ are appreciable.

\section{Conclusions}

In this talk we have presented a comparison between two  QCD
effective  charge,  $\alpha_{\mathrm{PT}}(q^2)$ and $\alpha_{\mathrm{gh}}(q^2)$, 
obtained from the PT and the  ghost-gluon  vertex, respectively.

Despite  their distinct theoretical origin, 
due  to a fundamental  identity relating the various ingredients entering into their definitions,  
the two effective charges are  almost identical in the entire range of
physical  momenta. In fact, the coincide  exactly in  the deep  infrared, where
they  freeze at  a common  finite value, signaling the appearance of IR 
fixed point in QCD~\cite{Brodsky:2003px}, also required
from a  variety of phenomenological studies~\cite{Halzen:1992vd}.

{\it Acknowledgments:} 
The authors thank the organizers of LC09 for their hospitality. 
The research of JP is supported by the European FEDER and  Spanish MICINN under grant FPA2008-02878, 
and the ``Fundaci\'on General'' of the University of Valencia.

\end{document}